\documentclass[prl,twocolumn,showpacs, preprintnumbers,superscriptaddress,amsmath,amssymb]{revtex4-1}
\usepackage{graphicx,color,bm,natbib, times, placeins} 
\usepackage[colorlinks=true,linkcolor=red,citecolor=green]{hyperref}
\usepackage{amssymb,amsfonts,amsmath,graphicx,amsthm}
\usepackage{color}

\begin{document}
\title{Data Fusion Reconstruction of Spatially Embedded Complex Networks}

\author{Jie Sun}
\email{sunj@clarkson.edu}
\affiliation{Clarkson Center for Complex Systems Science, Clarkson University, Potsdam, New York, 13699, USA} 
\affiliation{Department of Mathematics, Clarkson University, Potsdam, New York, 13699, USA} 
\affiliation{Department of Physics, Clarkson University, Potsdam, New York, 13699, USA} 
\affiliation{Department of Computer Science, Clarkson University, Potsdam, New York, 13699, USA} 
\author{Fernando J. Quevedo}
\affiliation{Clarkson Center for Complex Systems Science, Clarkson University, Potsdam, New York, 13699, USA} 
\affiliation{Department of Mathematics, Clarkson University, Potsdam, New York, 13699, USA} 
\affiliation{Department of Mechanical Engineering, Clarkson University, Potsdam, New York, 13699, USA} 
\author{Erik Bollt}
\affiliation{Clarkson Center for Complex Systems Science, Clarkson University, Potsdam, New York, 13699, USA} 
\affiliation{Department of Mathematics, Clarkson University, Potsdam, New York, 13699, USA} 
\affiliation{Department of Electrical \& Computer Engineering, Clarkson University, Potsdam, New York, 13699, USA}

%\date{\today}

\begin{abstract}%max 600 characters including space for PRL
We introduce a kernel Lasso (kLasso) optimization that simultaneously accounts for spatial regularity and network sparsity to reconstruct spatial complex networks from data. Through a kernel function, the proposed approach exploits spatial embedding distances to penalize overabundance of spatially long-distance connections.  Examples of both synthetic and real-world spatial networks show that the proposed method improves significantly upon existing network reconstruction techniques that mainly concerns sparsity but not spatial regularity.  Our results highlight the promise of data fusion in the reconstruction of complex networks, by utilizing both microscopic node-level dynamics (e.g., time series data) and macroscopic network-level information (metadata).
\end{abstract}

% pacs ref: https://www.aip.org/publishing/pacs/pacs-2010-regular-edition

\maketitle

%: Introduction
Reconstructing a complex network from observational data is an outstanding problem.  Successful network reconstruction can reveal important topological and dynamical features of a complex system and facilitate system design, prediction, and control, as demonstrated in several recent studies across multiple disciplines~\cite{Napoletani2008,Runge2012,Altarelli2014,Antulov-Fantulin2015,Runge2015,Ambegedara2016,Lord2017}. In many applications, such as material science, infrastructure engineering, neural sensing and processing, and transportation, the underlying complex networks are often spatially embedded (see~\cite{Barthelemy2011} for an excellent review). The spatial embedding adds yet another dimension to the problem of complex network reconstruction.

A common property of spatially embedded networks is {\it spatial regularity}, which manifests itself as an inverse dependence of connection probability on spatial distance: generally, the larger the spatial distance is between two nodes, the less likely there exists an edge connecting these nodes~\cite{Barthelemy2011}. This feature of spatial networks, which can be attributed to physical, financial, or other constraints, has been observed in various types of spatial networks from several different studies, including street patterns~\cite{Masucci2009}, mobile communication~\cite{Lambiotte2008}, and social networks~\cite{Liben-Nowell2005}. Indeed, the interdependence between network structure and spatial distance is a key ingredient in many widely used models and important studies of spatially embedded networks~\cite{Waxman1988,Kleinberg2000,Rozenfeld2002,Barrat2005,Gonzalez2006,Carmi2009,Bradde2010,Expert2011,Henderson2011,Frasco2014,Shandeepa2017}.

\begin{figure}[htbp]
\centering
\includegraphics[width=0.47\textwidth]{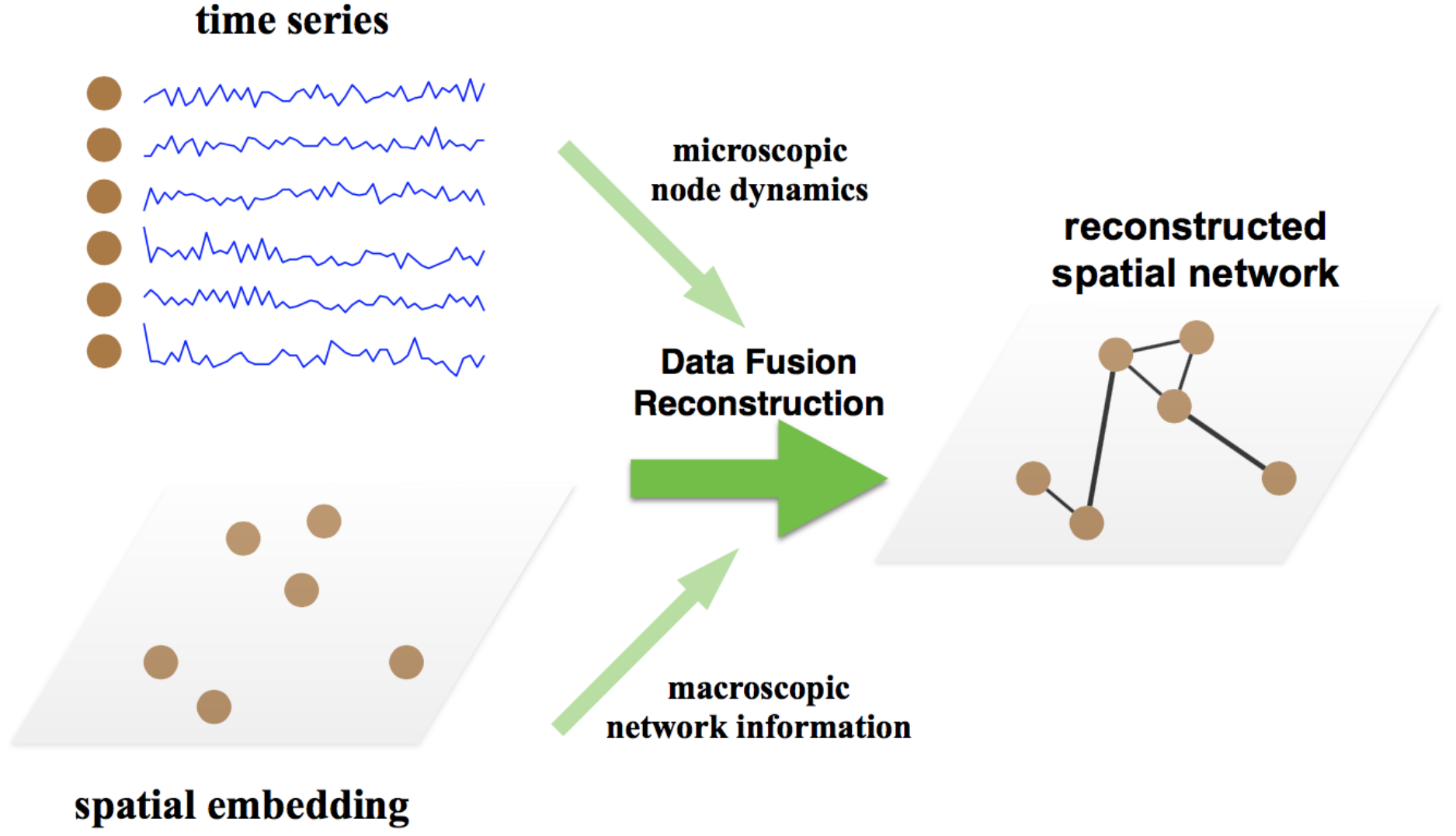} 
\caption{(color online.) Data fusion reconstruction involves the appropriate ``fusion" of information from various sources. For example, such information can arise from both microscopic node-level dynamics (such as the time series of the individual nodes) and macroscopic network-level meta data (e.g., spatial embedding of the nodes).}
\label{fig:1}
\end{figure}

In this Letter, we show that spatial regularity can be exploited to significantly enhance the accuracy of network reconstruction. In particular, in view of the often limited amount of data available for the inference of large complex networks, the central challenge has always been to better utilize information that potentially arise from distinct sources. To this end, we propose {\it data fusion reconstruction (DFR)} as a principal framework to infer networks in the presence of both microscopic dynamic data (e.g., time series) and metadata (i.e., spatial embedding information). See Fig.~\ref{fig:1}. To demonstrate the concept of DFR, we developed kernel Lasso (kLasso) as a generalization of the Lasso, the latter is widely used for sparse regression~\cite{Tibshirani1996}. Using examples of both synthetic and real-world spatial networks, we show that due to the integration of sparsity and spatial regularity effects, 
kLasso reconstructs spatial networks significantly better than Lasso.
%: FIGURE 1: cartoon figure %%%%%
%%%%% END OF FIGURE 1 %%%%%

Mathematically, a spatially embedded network can be represented by a triplet of sets, $G=(V;E;\Phi)$, where $V=\{1,2,\dots,n\}$ is the set of nodes, $E=\{(i_k,j_k)_{k=1}^{m}\}\subset V\times V$ is the set of (directed) edges, and $\Phi=[\mathbf{\Phi}_1,\mathbf{\Phi}_2,\dots,\mathbf{\Phi}_n]$ encodes the spatial embedding of the nodes: for example, for a $q$-dimensional Euclidean embedding, $\mathbf{\Phi}_i\in\mathbb{R}^q$. 

%: Problem setup
{\it Problem setup.} Given time series data as well as spatial location of the nodes, the problem is to reconstruct the underlying network structure. 
We represent the time series data as $\{x_j(t)\}_{j=1,\dots,n;t=0,1,\dots,T}$, where $x_j(t)$ denotes the observed state of node $j$ at the $t$-th time instance. 
The sample size $T$ is the number of times each node is observed. 
In addition, the spatial coordinates of the nodes give rise to an embedded distance $d_{ij}=d(\mathbf{\Phi}_i,\mathbf{\Phi}_j)$ defined for each pair of nodes $(i,j)$.

To represent the interactions among the nodes, we employ a standard time series modeling approach~\cite{Brockwell2005}, by seeking a (stochastic) linear dependence of the state of each node $i$ at time $t$ on the state of all nodes at time $t-1$:
\begin{equation}\label{eq_linear}
	x_i(t) = \sum_{j=1}^{n}A_{ij}x_j(t-1) + \xi_i(t),~\mbox{for~}i=1,2,\dots,n.
\end{equation}
The network structure is encoded in the adjacency matrix $A=[A_{ij}]_{n\times n}$, where $A_{ij}\neq0$ if the state of node $i$ depends on the state of node $j$, and $A_{ij}=0$ otherwise. The extra term $\xi_i(t)$ denotes (dynamical) noise.
In the case where both $x_i(0)$ and $\xi_i(t)$ are Gaussian, then so is $x_i(t)$, and the vector $\mathbf{x}(t)=[x_1(t),\dots,x_n(t)]^\top$ follows a multivariate Gaussian distribution. Hidden behind the deceptively simple form of Eq.~\eqref{eq_linear} is complexity encoded by the network structure as represented by matrix $A$, a key factor that enables such standard model to be applicable to broad research topics such as information coding and communication~\cite{Cover2006}, linearization of nonlinear dynamics~\cite{Lasota1994}, statistical learning~\cite{Hastie2015}, and as a fundamental model form underlying dynamic mode decomposition~\cite{Schmid2010} and Koopman analysis of nonlinear systems~\cite{Mezic2013,Williams2015}.

%: Kernel-Lasso formulation of data fusion reconstruction
{\it Kernel Lasso.} We breakdown the reconstruction problem into the inference of each node's set of neighbors, $\mathcal{N}_i=\{j:A_{ij}\neq0\}$. Given $i$, we define $\mathbf{b}=[x_i(1),x_i(2),\dots,x_i(T)]^\top$, $M=[M_{tj}]_{T\times n}$ where $M_{tj}=x_j(t-1)$, and  $\mathbf{z}=[A_{i1},\dots,A_{in}]^\top.$ We collect the spatial distance between $i$ and the other nodes into a vector $\mathbf{s}=[s_j]_{n\times1}$ with $s_j=d_{ij}$. We propose the kernel Lasso (kLasso) optimization problem
\begin{equation}\label{eq_kernel}
\underset{\mathbf{z}}{\min} \left( \|M\mathbf{z}-\mathbf{b}\|_2^2 + \lambda\langle\kappa(\mathbf{s}),|\mathbf{z}|\rangle \right),
\end{equation}
where $|\mathbf{z}|=[|z_1|,\dots,|z_n|]^\top$, $\langle\cdot,\cdot\rangle$ denotes inner product, and $\kappa(\mathbf{s})=[\kappa(s_1),\dots,\kappa(s_n)]^\top$ is obtained by applying a scalar-valued kernel function $\kappa(\cdot)$ to the spatial distances to facilitate preference of spatially short-distance edges over long-distance ones. Finally, the regularization parameter $\lambda\geq0$ controls the tradeoff between model fit and model regularity. kLasso generalizes the classical Lasso formulation: when the kernel function is a constant, kLasso reduces to Lasso. As we demonstrate later using both synthetic and real-world spatial networks, by explicitly account for spatial embedding information, kLasso generally achieves better reconstruction.

Next we show how to solve kLasso problems. Consider an arbitrary kernel function $\kappa:\mathbb{R}\rightarrow\mathbb{R}^+$. Define matrix $\tilde{M}=[\tilde{M}_{tj}]_{T\times n}$ and vector $\tilde{\mathbf{z}}=[\tilde{z}_1,\dots,\tilde{z}_n]^\top$ as follows:
\begin{equation}
\begin{cases}
\tilde{M}_{tj}=M_{tj}/\kappa(s_j),\\
\tilde{z}_j=\kappa(s_j)z_j.
\end{cases}
\end{equation}
Applying these transformations to Eq.~\eqref{eq_kernel} converts a kLasso problem into a Lasso problem:
$\underset{\tilde{\mathbf{z}}}{\min} \left( \|\tilde{M}\tilde{\mathbf{z}}-\mathbf{b}\|_2^2 + \lambda\|\tilde{\mathbf{z}}\|_1 \right)$, which can be efficiently solved using standard algorithms (such as sequential least squares) found in the literature of computational inverse problems and statistics~\cite{Hastie2015}.

Here we focus on a general class of kernel functions of the shifted power-law form 
\begin{equation}\label{eq_kappa}
\kappa(d)=(d+d_0)^{\gamma},
\end{equation}
where the parameter $d_0>0$ ensures that $\kappa(d)>0$ for all $d\geq0$ whenever $\gamma\geq0$. On the other hand, the kernel exponent $\gamma\geq0$ is used to tune the preference toward spatial regularity: the choice of $\gamma=0$ recovers the Lasso solution, while the other extreme of $\gamma\rightarrow\infty$ ``selects" only the edges that have shortest spatial distance to each node. Intermediate values of $\gamma$ typically result in a more balanced mix of short-distance edges and long-distance edges appearing in the reconstructed network. Unless otherwise noted, we set the parameters at the default values $\gamma=1$ and $d_0=\min_{i\neq j}d_{ij}>0$.

%: Synthetic network example
{\it Synthetic network example: data-enabled inference of random spatial networks.} To benchmark the proposed kLasso method, we consider random spatial networks generated by the Waxman model~\cite{Waxman1988}. 
In particular, for each node pair $(i,j)$ whose spatial distance is $d_{ij}$, the probability of having an edge between $i$ and $j$ follows
\begin{equation}\label{eq_Waxman}
	P(d_{ij})= ce^{-\alpha d_{ij}},
\end{equation}
where $c>0$, and $\alpha\geq0$ (the special case of $\alpha=0$ produces a classical Erd\H{o}s-R\'{e}nyi random network embedded in space) with larger values of $\alpha$ lead to relatively more short-distance edges as compared to long-distance edges. 
For fixed $\alpha$, larger values of $c$ generally result in denser networks.
%: FIGURE 2: Waxman model %%%%%
\begin{figure}[htbp]
\centering
\includegraphics[width=0.5\textwidth]{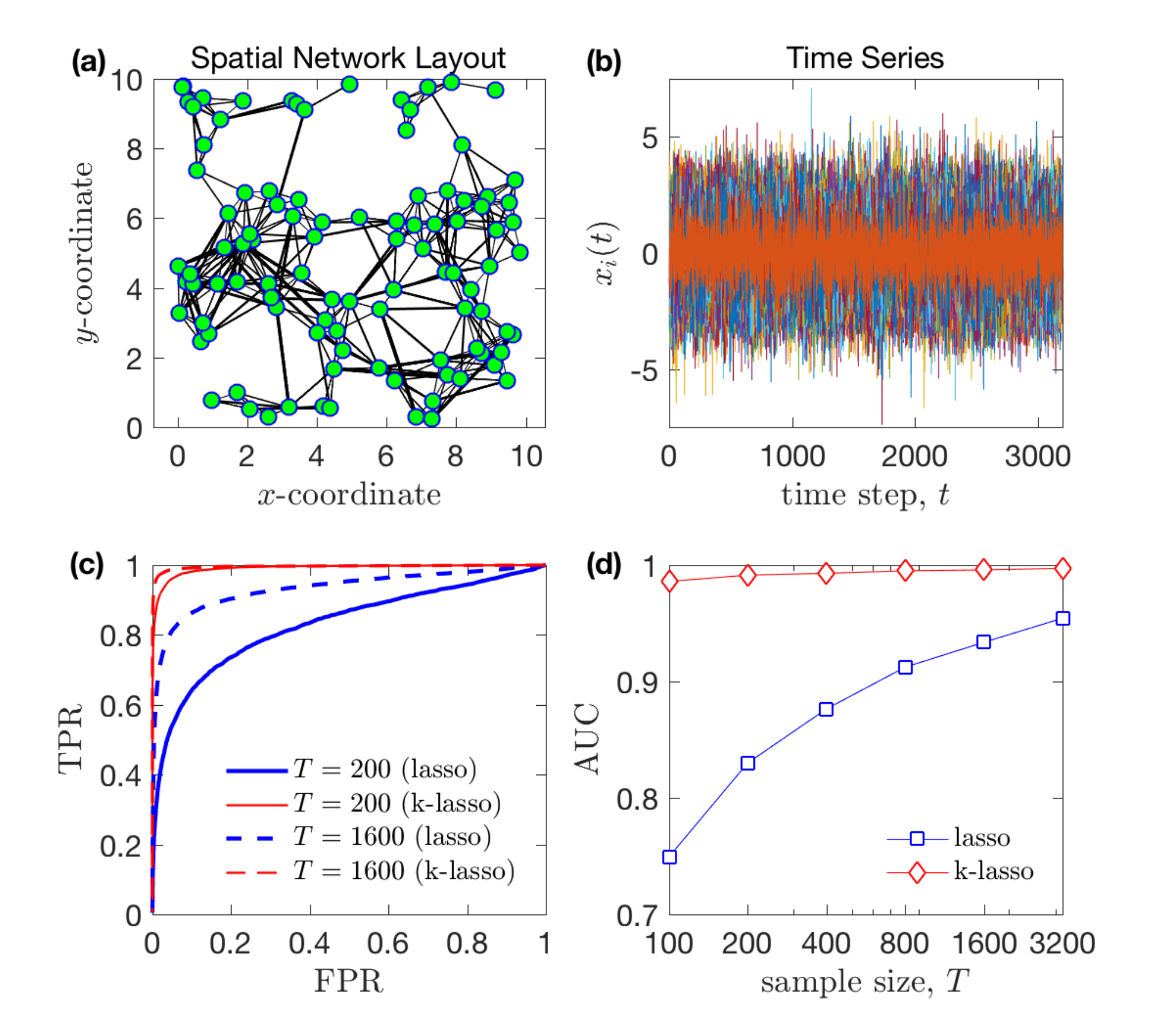} 
\caption{(color online.) (a) Layout of a random spatial network of $n=100$ nodes.
(b) Typical time series obtained by the stochastic dynamics on the network (see main text for details).
(c) Quality of network reconstruction as shown by ROC curves using Lasso versus using kLasso, for two sample sizes.
(d) Additional comparison between Lasso and kLasso in network reconstruction based on AUC values across a range of sample sizes.
Each data point in (c) and (d) represents an average over $10$ independent numerical experiments.
}
\label{fig:2}
\end{figure}
%%%%% END OF FIGURE 2 %%%%%

We show an example spatial network in Fig.~\ref{fig:2}(a). The network contains $n=100$ nodes that are randomly placed in the 2D spatial domain $(0,\sqrt{n})^2$. The structure of the network is generated according to Eq.~\eqref{eq_Waxman} with $\alpha=2$ and $c=10$, resulting in a total of $m=774$ edges. In addition, we generate a self-loop at each node to resemble short-term memory effects. For each edge $(i,j)$ (including the self loops), the weights $A_{ij}$ and $A_{ji}$ are independently drawn from the uniform distribution in $[-1,1]$. After that, the entire matrix $A$ is scaled $A\rightarrow cA$ with some constant $c$ such that $\rho(A)$, the spectral radius of $A$, is smaller than 1 to ensure stability of the stochastic process. We select $c$ to yield $\rho(A)=0.9$.
Stochastic time series data is obtained from the network dynamics~\eqref{eq_linear} using iid Gaussian noise $\xi_i(t)\sim\mathcal{N}(0,1)$. After discarding initial transient, data from  $T$ time steps is used for reconstruction. Typical time series of the network is shown in Fig.~\ref{fig:2}(b).

We compare the results of network reconstruction using kLasso [Eq.~\eqref{eq_kernel} with $\gamma=3$] versus Lasso. To measure the quality of reconstruction, we compute, for each estimate $\hat{A}$ of $A$, the true positive rate (TPR) and false positive rate (FPR) as
%\begin{equation}
\[\begin{cases}
\mbox{TPR}=|\{(i,j):\widehat{A}_{ij}\neq0~\&~A_{ij}\neq0\}|/|\{(i,j):A_{ij}\neq0\}|,\\
\mbox{FPR}=|\{(i,j):\widehat{A}_{ij}\neq0~\&~A_{ij}=0\}|/|\{(i,j):A_{ij}=0\}|.
\end{cases}\]
%\end{equation}
In Fig.~\ref{fig:2}(c) we plot the receiver operating characteristic (ROC) curves resulted from kLasso versus Lasso. An ROC curve shows the relationship between TPR and FPR as the regularization parameter $\lambda$ is varied. Exact, error-free reconstruction corresponds to the upper-left corner of the unit square $[0,1]^2$ ($\mbox{TPR}=1$, $\mbox{FPR}=0$), whereas reconstruction by random guesses would yield a diagonal line connecting $(0,0)$ (empty network) to $(1,1)$ (complete network). Each ROC curve can be summarized by a scalar defined as the area under the curve (AUC). AUC values are bounded between $0$ and $1$, with the larger the AUC, generally the closer the ROC curve is to the upper-left error-free corner and the better the reconstruction (AUC value of $1$ corresponds to exact reconstruction). As shown in Fig.~\ref{fig:2}(d) for a wide range of sample sizes, kLasso yields significant improvement over Lasso for network reconstruction.
The key reason behind kLasso's success in reconstructing spatial networks lies in its unique capability to incorporate spatial embedding information to ``penalize" formation of edges that span over larger spatial distances.

%: Application
{\it Application: reconstruction from hidden individual dynamics.} We now turn to an application of reconstructing a transportation network from observable population-level dynamics data that result from hidden individual trajectories. 

\begin{figure}[tbp]
\centering
\includegraphics[width=0.38\textwidth]{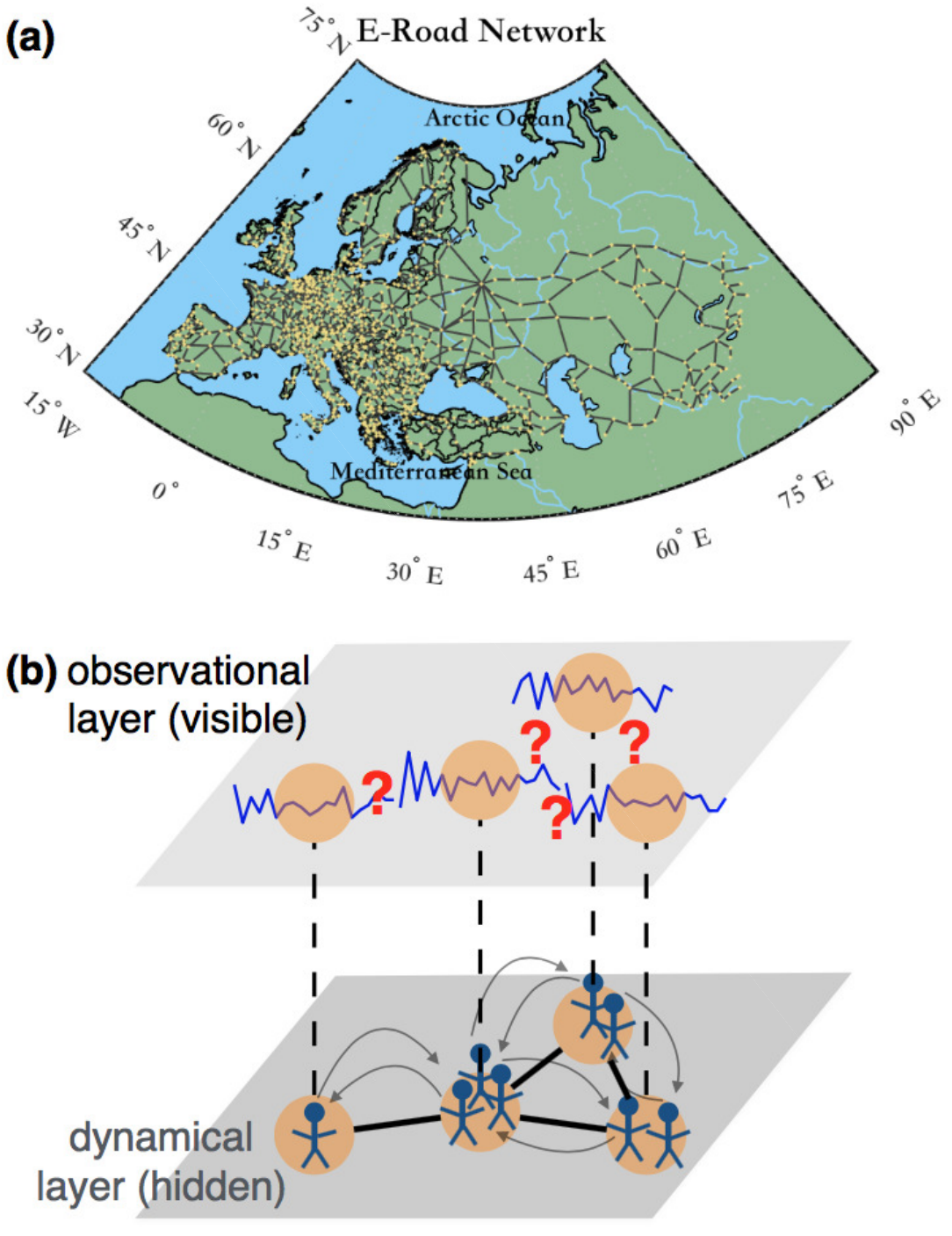} 
\caption{(color online.) (a) Visualization of the E-Road network, where nodes (marked by light yellow dots) correspond to cities in the Europe and edges (marked by dark gray lines) represent highway segments connecting the cities (water crossings are excluded). This spatial network contains  $n=955$ nodes and $m=1255$ edges.
(b) Two-layer illustration of the hidden dynamics on hidden network, where both the network structure and the dynamics of individuals on the network are hidden (hidden dynamical layer), only the aggregated population dynamics is measured (observational layer).
}
\label{fig:3}
\end{figure}

The network here is a continent-scale transportation network of Europe, referred to as the E-Road network~\cite{ERoad}, visualized in Fig.~\ref{fig:3}(a). A node represents a city of Europe whereas an edge between two nodes represents a highway segment that directly connects the corresponding cities. We compute the embedding distance between each pair of nodes as the shortest distance along the Earth's surface using the corresponding cities' latitude and longitude information. 

The dynamical system here describes hidden dynamics on a hidden network, and can be conceptually understood by considering two layers, as illustrated in Fig.~\ref{fig:3}(b). On the hidden, dynamical layer, there is a total of $N$ individuals, each moving around independently in the spatial network by following a discrete-time random walk~\cite{Noh2004}. At each time step, an individual at node $i$ moves along one of the edges $(i,j)$ in the network at random to reach node $j$. On the observational layer, the {\it aggregated} number of individuals at each node is observed, producing a time series $\{x_i(t)\}$, where $x_i(t)$ is the aggregated number of individual walkers occupying node $i$ at time $t$. 
The problem is to reconstruct the hidden spatial network from the observed time series of the population dynamics in the absence of individual trajectories.
%: FIGURE 3: E-Road network and population dynamics%%%%%
%%%%% END OF FIGURE 3 %%%%%

\begin{figure}[tbp]
\centering
\includegraphics[width=0.37\textwidth]{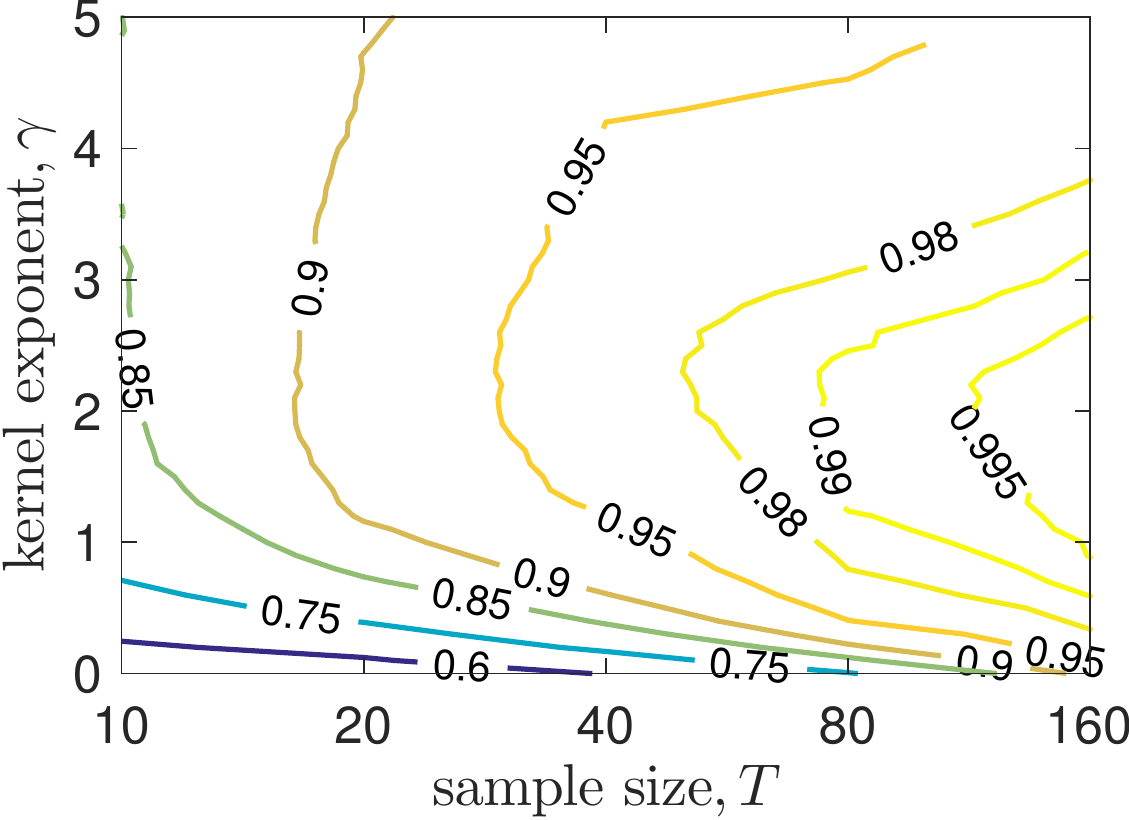} 
\caption{(color online.) 
Data fusion reconstruction of the E-Road network using kLasso, shown as a contour plot of AUC values in the $(T,\gamma)$ plane. Note that for fixed sample size $T$, Lasso corresponds to choosing $\gamma=0$, which always yields lower AUC (worse reconstruction) than using kLasso ($\gamma>0$). Here the optimal choice of $\gamma$ lies somewhere between $2$ and $3$.
}
\label{fig:4}
\end{figure}

Figure~\ref{fig:4} shows network reconstruction results using kLasso across a range of kernel exponents $\gamma$ and sample size $T$. 
Excellent reconstruction is generally achieved, with AUC value starts to increase above $0.99$ for sample size as low as $T\approx80$, a number that is surprisingly small compared to the size of the network ($n=955$ nodes, $m=1255$ edges). kLasso better reconstructs the network than Lasso (corresponds to $\gamma=0$) in all parameter combinations, with most significant increase of AUC occurring for $1\lesssim\gamma\lesssim3$. For fixed $\gamma$, improvement is more significant for smaller $T$.
In addition to further validating the effectiveness of kLasso, the example demonstrates the possibility to reconstruct a hidden spatial network by merely observing aggregated population-level dynamics instead of having to following detailed individual trajectories.

%: FIGURE 4: E-road inference results %%%%%
%%%%% END OF FIGURE 4 %%%%%

%: Conclusion
To summarize, we here developed a kLasso approach for data fusion reconstruction of spatially embedded complex networks. We show that under appropriate linear transformations, kLasso can be converted into a corresponding Lasso problem and thus be efficiently solved using standard Lasso algorithms. We benchmark the effectiveness of kLasso using data from stochastic dynamics on a random spatial network. Furthermore, we consider hidden individual dynamics on E-Road network (a real-world transportation network) where the only observables are the aggregated population dynamics over spatially embedded node locations. kLasso attains excellent reconstruction of the network without the need to fine-tune parameters even for very short time series. 
These results demonstrates the power of data fusion in the inference in complex systems, in particular the utility of kLasso in the efficient and effective reconstruction of spatially embedded complex networks, when there is both microscopic (e.g., time series data on the nodes) and macroscopic (e.g., metadata of the network) information. Given the flexibility of designing the kernel, it will be interesting to explore other types of metadata for enhanced network resonstruction, such as occupation in social networks.
Reconstruction of the E-Road network despite unobservable individual dynamics suggests the possibility of inferring transportation channels from population-level observations without the necessity to trace detailed individual trajectories. This makes kLasso a potentially useful tool for uncovering hidden spatial mobility patterns in practice.

%\subsection{Acknowledgments}
%: acknowledgments
\acknowledgments
This work was funded in part by the Army Research Office grant W911NF-16-1-0081, the Simons Foundation grant 318812, the Office of Naval Research grant N00014-15-1-2093, and the Clarkson University Space Grant Program an affiliate of the New York NASA Space Grant Consortium.

%: References

\end{document}